\documentclass{jps-cp}
\usepackage{txfonts} %Please comment out this line unless the txfonts package is availabe in your LaTeX system.
\usepackage{graphicx}
\usepackage{xcolor}

\usepackage{amsmath}
\title{Embedding and correlation tensor for XRP transaction networks}

\author{Abhijit \textsc{Chakraborty}$^{1,2,*}$, Tetsuo \textsc{Hatsuda}$^{2}$ and Yuichi \textsc{Ikeda}$^{1}$}

\inst{$^{1}$ Kyoto University, Graduate School of Advanced Integrated Studies in Human Survivability, Kyoto, 606-8306, Japan \\
$^{2}$RIKEN Interdisciplinary Theoretical and Mathematical Sciences Program, Saitama, 351-0198, Japan}

\email{*chakraborty.abhijit.7y@kyoto-u.ac.jp}

\recdate{October 1, 2022}

\abst{
Cryptoassets are growing rapidly worldwide. One of the large cap cryptoassets is XRP. In this article, we focus on analyzing transaction data for the 2017–2018 period that consist one of the significant XRP market price bursts. 
We construct weekly weighted directed networks of XRP transactions.  These weekly networks are embedded on continuous vector space using a network embedding technique that encodes structural regularities present in the network
structure in terms of node vectors. Using a suitable time window we calculate a correlation tensor. A double singular value decomposition of the correlation tensor provides key insights about the system. The
significance of the correlation tensor is captured using a randomized correlation tensor. We present a detailed dependence of correlation tensor on model parameters. 
}

\kword{cryptoaasets, XRP, network embedding, correlation tensor, double singular value
decomposition}

\begin{document}
\maketitle

\section{Introduction}
Cryptoassets are digital assets, which use cryptography and depend on distributed ledger technology, also known as blockchain. The cryptomarket consists of many cryptoassets, namely, Bitcoin, Etherium, and XRP etc. Recently, we have witnessed increase in investor attention on cryptoassets.
However, the risk for investing in cryptoassets is generally high due to the high volatility of their market price. Moreover, as the growing cryptoassets market can disrupt the financial market and the possibility of money laundering, policymakers 
in different countries have already started enacting regulations for its use. The cryptoassets have also attracted attention from researchers of different fields to study the fluctuation in price, transaction patterns of users, and  efficiency, robustness of the underlying blockchain technology.          
While there are many studies on Bitcoin and Etherium~\cite{wu2021analysis, kondor2014do, ferretti2020onthe}, there are very few studies on XRP~\cite{ikeda2022characterization, aoyama2022Cryptoasset}. In this work, we focus on analyzing XRP transactions.

Time series analysis comprises different methods to characterize and extract crucial insights for various time series data in the stock markets~\cite{laloux1999noise, plerou1999universal, plerou2002random}, foreign exchange markets~\cite{chakraborty2018deviations, chakraborty2020uncovering } or even in medical recordings~\cite{schindler2007assessing}. The cross correlation is one of the popular methods to analyze time series data~\cite{laloux1999noise, plerou1999universal, plerou2002random, schindler2007assessing}. 
The simplest way to measure correlation is the Pearson correlation, which is defined for a pair of variables $(x,y)$ as $r_{x,y} = \sum\limits_{i=1}^n (x_i-\overline{x})(y_i-\overline{y})/((n-1)\sigma_x \sigma_y)$. Here $n$ is the no of observations or the length of the time series. The mean and standard deviations of $x$ and $y$ are represented as $\overline{x}, \overline{y}$, and $\sigma_x, \sigma_y$. The applications of cross correlation method with random matrix theory on time series data have provided valuable insights for different systems. 

Recently, inspired by the cross correlation analysis of time series, A. Chakraborty {\em et al.} have developed a method of correlation tensor spectra for dynamical XRP transaction networks~\cite{chakraborty2022projecting} to capture price burst. Following this article, we show the details of the embedding method for weekly directed weighted XRP networks. Using the embedded node vectors, we calculate the correlation tensors for XRP transaction netwokrs. The significance of the elements of the correlation tensor and its spectra is shown by comparing with randomized correlation tensor. Furthermore, We uncover the dependence of the correlation tensor on various model parameters, such as the embedding dimension, and the time window in detail.

\section{Data description}
Our data consist of the direct transactions between different XRP wallets from October 2, 2017 to March 4, 2018.  The dataset was recorded as ledger data using the Ripple Transaction Protocol.
We divided this dataset into $22$ groups based on different weeks. We constructed a weekly directed weighted network of XRP transactions by aggregating all the transactions for a week. In this network, XRP wallets are the nodes and a link represents the flow of XRP from a source wallet to a destination wallet. 
The total amount of weekly XRP flow between a pair of wallets represents the link weight.  
%See~\cite{ikeda2022characterization} for the structural properties of the XRP transaction network. 
The XRP transactions data were collected from the ripple data API at~{https://xrpl.org/data-api.html}.

%\section{Methods}
\section{Embedding of XRP transaction networks}
%intro to embedding
Network embedding is a technique for representation of a network in low dimensional vector space by preserving key network features, which is useful for downstream tasks, such as network visualization, network analysis, link predication etc. 
The DeepWalk~\cite{perozzi2014deepwalk} and node2vec~\cite{grover2016node2vec} are the two well-known methods for network embedding with structural information. We briefly describe these two methods below. 

\subsection{DeepWalk}
Following the natural language models~\cite{collobert2008unified}, B. Perozzi {\em et al.} developed  the DeepWalk algorithm~\cite{perozzi2014deepwalk}, which maps each node $v_i \in V$ of a network as a vector in a $D$ dimensional continuous space $R^{{\mid V \mid} \times D}$  by modeling a set of truncated 
random walks. The algorithm encodes the network community structure into a vector representation of the nodes. 
Many short random walk are used to compute local community structure information efficiently. These walks correspond to short sentences in language modeling.
More specifically, it uses many random walks $R_w$ of length $l$ from each node. For each node, the algorithm generates a random walk of length $l$ and uses it to update the SkipGram algorithm in accordance with an objective function. The objective function maximizes the co-occurrence probability with other nodes present in the short random walk sequence. For a mapping function $\phi: v \in V \to R^{{\mid V \mid} \times D}$, the objective function can be written as 
\begin{equation}
\min_{\phi} -\log Pr \biggl(\{v_{i-w},...,v_{i+w}\} \backslash v_i|\phi(v_i)\biggr),
\end{equation}
where $w$ is the window size, $\{v_{i-w},...,v_{i+w}\}$ is the node set generated by the random walks from $v_i$ and $\phi(v_i)$ represents the $v_i$ on the vector space. The algorithm uses the hierarchical softmax~\cite{{Minh2009scaleable}} to speed up the optimization procedure.

\subsection{node2vec}
A. Grover {\em et al.} further generalized the Deepwalk method by modifying the random walks as second order biased random walks. This method can encode more complex regularities of the network, such as functional relationships into the vector representation of nodes, depending on the tuning parameters $p$ and $q$ for the walks . It becomes equivalent to DeepWalk method when $p = q = 1 $. 

Here we have used node2vec with $p = q = 1 $ to embed weighted directed weekly XRP transactions networks.  

%\subsubsection{Subsubsection}

\section{Correlation tensor from embedded XRP transaction networks}
%---------------Figure 1-----------------
\begin{figure}[tbh]
\centering
\includegraphics[width=0.98\textwidth]{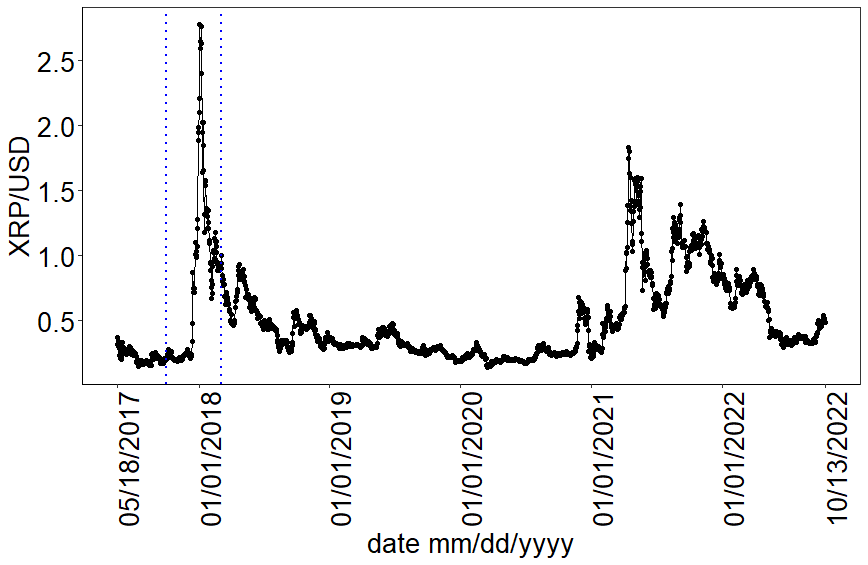}
\caption{The daily XRP/USD close price from May 05, 2017 to October 13, 2022. The dotted blue lines indicate the period that we consider for our analysis. 
This figure is adapted from~\cite{chakraborty2022projecting}.
%You can embed figures using the \texttt{\textbackslash includegraphics} command. Basically, figures should appear where they are cited in the text. You do not need to separate figures from the main text when you use \LaTeX\ for preparing your manuscript.
}
\label{f1}
\end{figure}
%--------------------------
%\begin{table}[tbh]
%\caption{Captions to tables and figures should be sentences.}
%\label{t1}
%\begin{tabular}{ll}
%\hline
%AAA & BBB \\
%CCC & DDD \\
%\hline
%\end{tabular}
%\end{table}

%\subsubsection{Equation numbers}

%The \verb|seceq| option resets the equation numbers at the start of each section.

We embed each of these $22$ weeekly weighted directed networks using node2vec with $p = q = 1 $.  It gives vector representations for each node $V_i^\alpha (t)$, where $t=1,2,3,\cdots,22$; $\alpha =1, 2, 3,\cdots, D$; $i =1, 2, 3,\cdots, N_t$. Here $D$ represents the dimension of the embedding space and $N_t$    
is the total number of nodes in the $t$-th week network. We identify that there are $N=71$ nodes that carry out at least one transaction each week. We call these $N=71$ nodes, regular nodes. The regular nodes play key role in the XRP trading market as they carry out frequent and consistent transactions.
Considering the embedded vectors of regular nodes $V_i^\alpha (t)$, where $i =1, 2, 3,\cdots, N$, we calculate the correlation tensor as follows 

\begin{equation}
M_{ij}^{\alpha\beta}(t) =\frac{1}{2\Delta T}\sum\limits_{t^\prime=t-\Delta T}^{t+\Delta T}\frac{[V_{i}^\alpha (t^\prime) - \overline{V_{i}^\alpha}][V_{j}^\beta (t^\prime) - \overline{ V_{j}^\beta}]}{\sigma_{V_i^\alpha} \sigma_{V_j^\beta}},  
\label{eqn1}
\end{equation}
where $\sum$ is taken over $(2\Delta T+1)$ weekly $\{t-\Delta T, t-(\Delta T-1),\cdots, t,\cdots, t+(\Delta T-1), t+\Delta T \}$ networks. The $ \overline{V_{i}^\alpha} $ and $ \sigma_{V_i^\alpha}$ represents mean and standard deviation of $V_{i}^\alpha$ over a time window of $(2 \Delta T+ 1)$  networks.

The correlation tensor has $(N \times N \times D \times D)$ elements. To uncover the crucial insights from the correlation tensor, we use a double singular value decomposition (SVD) technique as follows: 
%We conduct the diagonalization of $M_{ij}^{\alpha\beta}$ in terms of $(ij)$-index and $(\alpha\beta)$-index successively by the bi-unitary transformation or equivalently SVD. 

As a first step, we conduct the diagonalization of $M_{ij}^{\alpha\beta}$ in terms of $(ij)$-index
\begin{equation}
M_{ij}^{\alpha\beta} = \sum\limits_{k=1}^N L_{ik}\sigma_k^{\alpha\beta} R_{kj},
\label{eqn2}
\end{equation}
and as a second step, we perform the digonalization in terms of $(\alpha\beta)$-index successively
\begin{equation}
\sigma_k^{\alpha\beta} = \sum\limits_{\gamma=1}^D \mathcal{L}^{\alpha\gamma} \rho_k^\gamma \mathcal{R}^{\gamma\beta}.
\label{eqn3}
\end{equation}
Then, using Eq.~\ref{eqn2} and  Eq.~\ref{eqn3} we have
\begin{equation}
M_{ij}^{\alpha\beta} = \sum\limits_{k=1}^N \sum\limits_{\gamma=1}^D \rho_k^\gamma (L_{ik} R_{kj}) (\mathcal{L}^{\alpha\gamma} \mathcal{R}^{\gamma\beta}). 
\label{eqn4}
\end{equation}
Here $ \rho_k^\gamma$ represents the $N \times D$ generalized real and positive singular values as the correlation tensor $M$ is real.

\section{Results}

%---------------Figure 2------------------
\begin{figure}[tbh]
\centering
\includegraphics[width=0.98\textwidth]{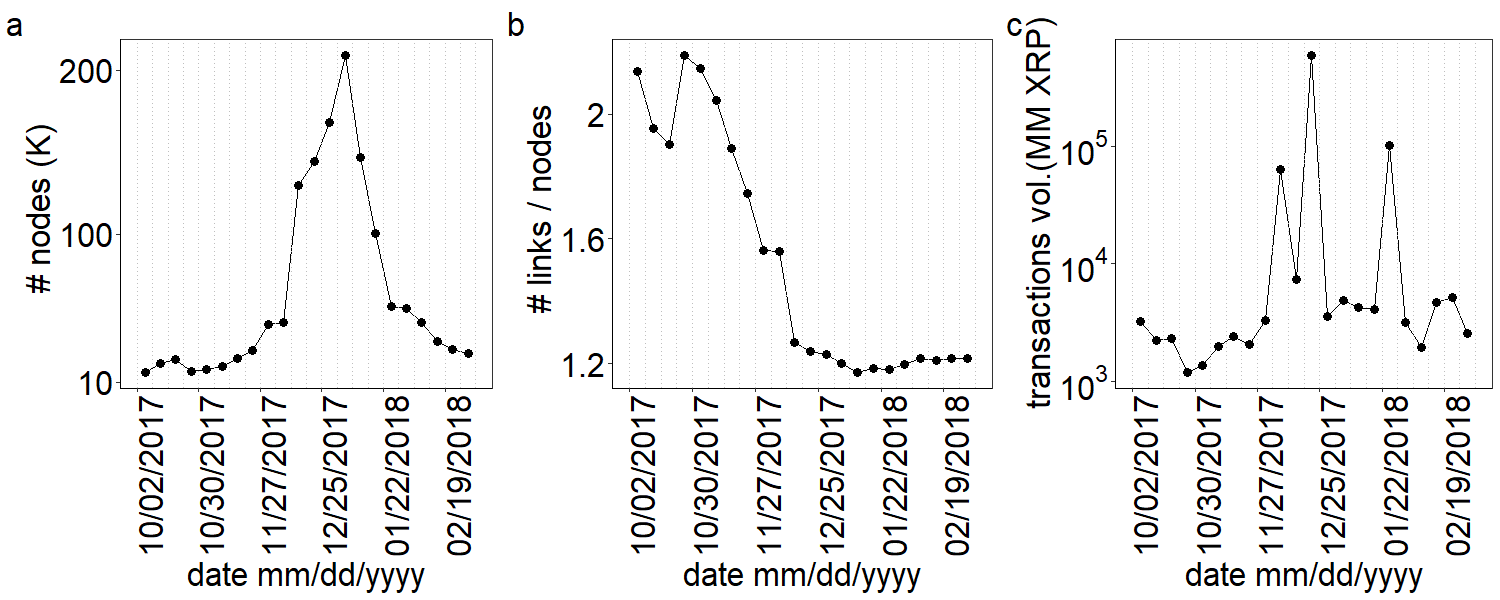}
\caption{ The temporal variation in the properties of directed weighted weekly XRP networks. Temporal variation of (a) node numbers, (b) number of links per node  and (c) total transaction volume. This figure is adapted from~\cite{chakraborty2022projecting}.
%You can embed figures using the \texttt{\textbackslash includegraphics} command. Basically, figures should appear where they are cited in the text. You do not need to separate figures from the main text when you use \LaTeX\ for preparing your manuscript.
}
\label{f2}
\end{figure}
%---------------------------------

%---------------Figure 3------------------
\begin{figure}[tbh]
\centering
\includegraphics[width=0.6\textwidth]{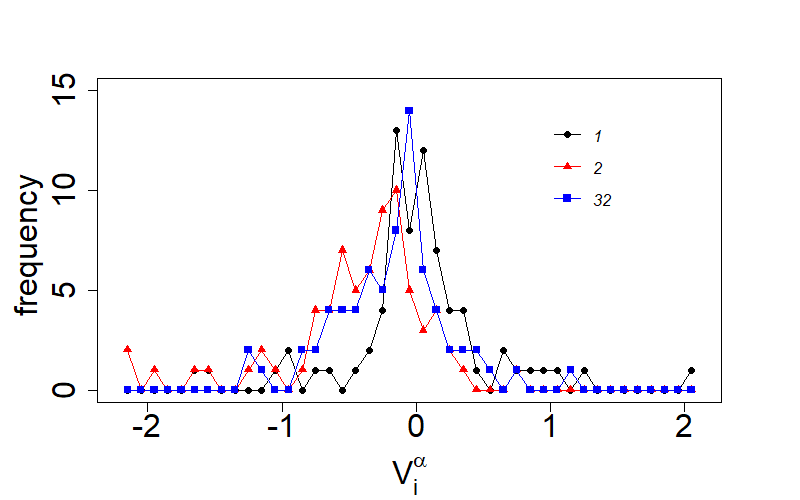}
\caption{Distributions for three different components $( \alpha = 1, 2,32$ are indicated as legends$ )$ of all regular nodes during the week, 2017, October 02 to October 08.     
%You can embed figures using the \texttt{\textbackslash includegraphics} command. Basically, figures should appear where they are cited in the text. You do not need to separate figures from the main text when you use \LaTeX\ for preparing your manuscript.
}
\label{f3}
\end{figure}
%---------------------------------
%----fig 1
We show the daily XRP/USD price between 2017, May 18 and 2022 October 13 in Fig.~\ref{f1}. The period consists several peaks of XRP/USD price. The most extraordinary rise and fall of XRP/USD price is observed around January 2018. 
So we focus our study between October 2, 2017 and March 4, 2018, which covers 22 weeks. 

We construct a weekly network of XRP transactions by aggregating all the transactions among the wallets. Wallets are the node and a directed link is formed between a source wallet and a destination wallet. The total flow of XRP between two wallets represents the link-weight. 
We show the temporal variation of number of nodes, links and value of total link-weight in Fig.~\ref{f2}. We observe that the number of nodes in the weekly networks increases rapidly around December 2017 and started decreasing again from the second week of January 2018, as shown in Fig.~\ref{f2}~(a).
Fig.~\ref{f2}~(b) shows that the number of links per node, also known as average in-degree or out-degree decreases from $2$ to $1.2$ during October, 2017-March 04, 2018. The variation of the total transaction volume which represents total link-weight in weekly networks, is shown in Fig.~\ref{f2}~(c). We observe
three sudden peaks in the transactions volume, which may be associated with the XRP price bubble during January 2018. 

The distributions for the components of the node vector $V_i^\alpha$ for all regular nodes for three different values of $\alpha$ are shown in  Fig.~\ref{f3} for the week 2017, October 02 to October 08. The distributions have a peak indicating that the nodes are not embedded randomly on the vector space but capture the regularities of the network. 

\begin{figure}[tbh]
%\centering
\includegraphics[width=0.98\textwidth]{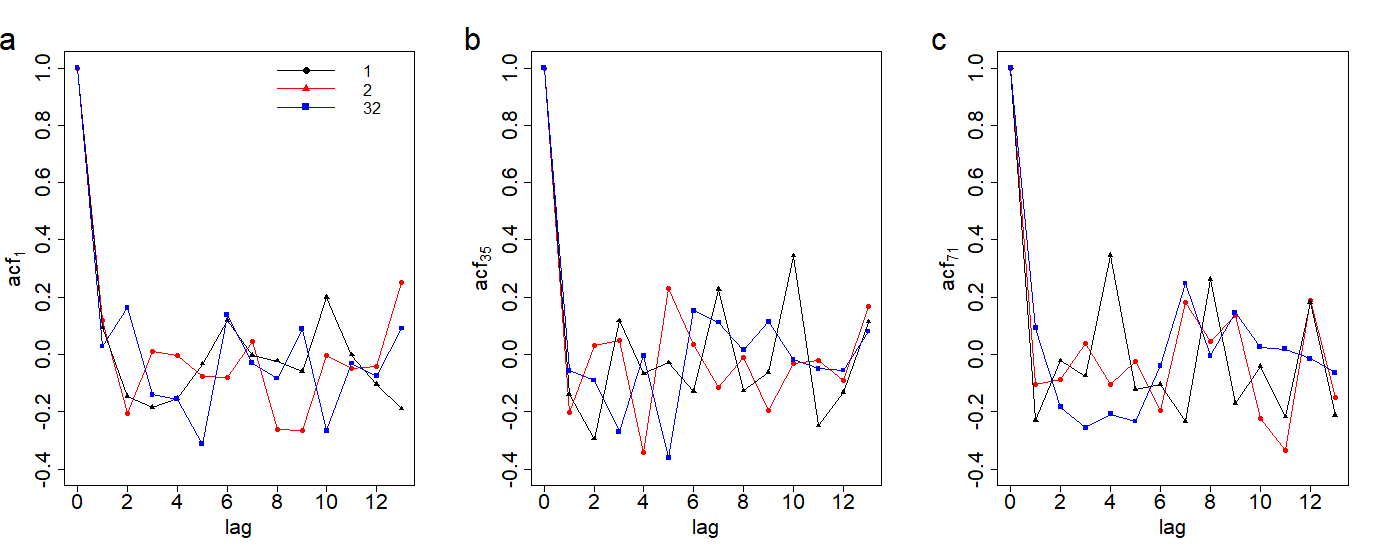}
\caption{The auto-correlation function acf$_i$ for three different components $( \alpha = 1, 2,32$ are indicated as legends$ )$ of three nodes $ i = {\rm (a)} ~1, {\rm (b)} ~35, {\rm and~ (c)} ~71$.     
%You can embed figures using the \texttt{\textbackslash includegraphics} command. Basically, figures should appear where they are cited in the text. You do not need to separate figures from the main text when you use \LaTeX\ for preparing your manuscript.
}
\label{f4}
\end{figure}

\begin{figure}[tbh]
%\centering
\includegraphics[width=0.98\textwidth]{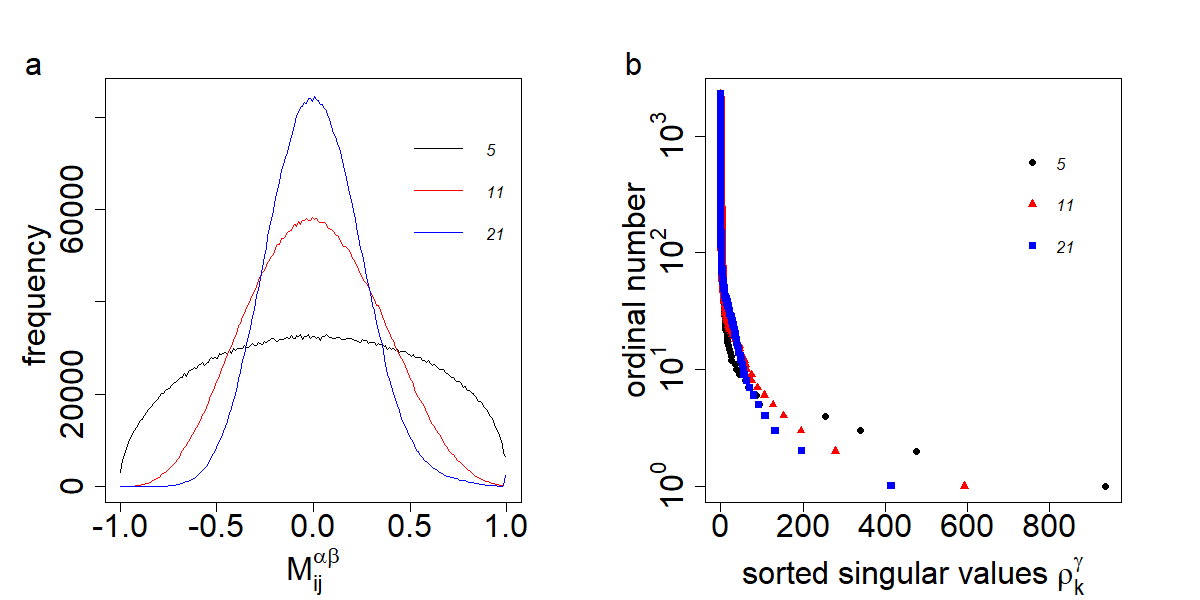}
\caption{(a) Distribution for the elements of correlation tensors $M(t)$ with different time windows $ 2\Delta T +1 = 5, 11$ and $21$, respectively. (b) Sorted singular values for the corresponding correlation tensors. 
The correlation tensors are calculated for the week, $t$ = 2017, October 16 - October 22, 2017, November 06 - November 12, and 2017, December 11 - December 17 respectively. 
%You can embed figures using the \texttt{\textbackslash includegraphics} command. Basically, figures should appear where they are cited in the text. You do not need to separate figures from the main text when you use \LaTeX\ for preparing your manuscript.
}
\label{f5}
\end{figure}
%sd(M5)
%[1] 0.5047187
%> sd(M11)
%[1] 0.3311724
%> sd(M21)
%[1] 0.2418685

\begin{figure}[tbh]
%\centering
\includegraphics[width=0.98\textwidth]{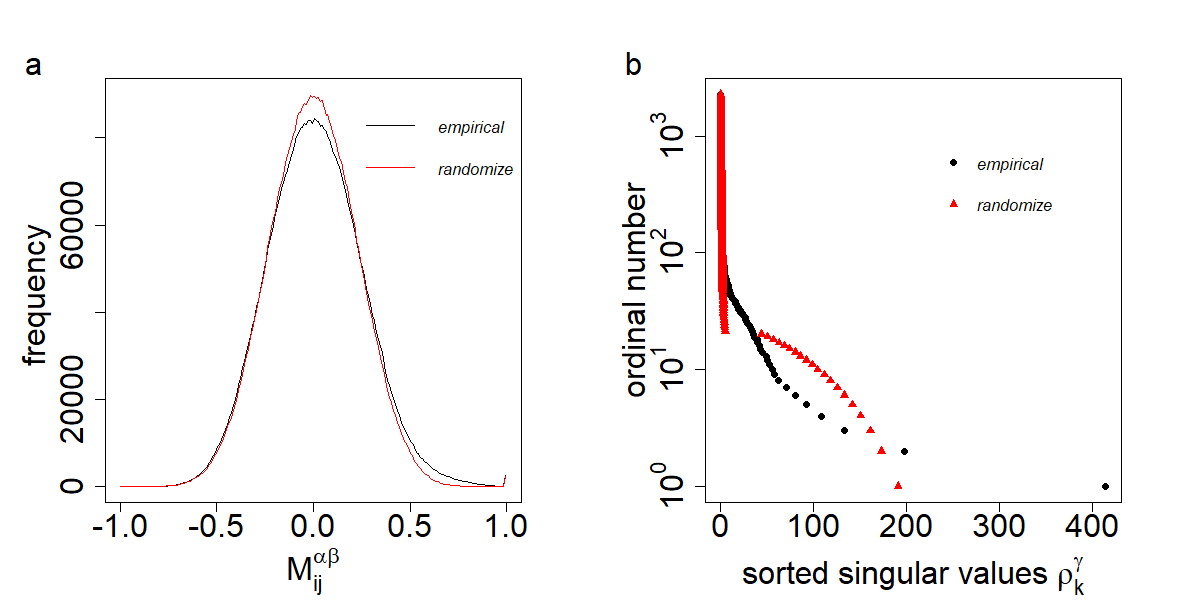}
\caption{Comparison between empirical correlation tensor and randomize correlation tensor with time window $2 \Delta T +1 = 21$.  (a) Distribution for the elements of correlation tensors.  (b) Sorted singular values for the corresponding correlation tensors. 
The correlation tensor is calculated for the week, $t$ = 2017, December 11 - December 17. 
%You can embed figures using the \texttt{\textbackslash includegraphics} command. Basically, figures should appear where they are cited in the text. You do not need to separate figures from the main text when you use \LaTeX\ for preparing your manuscript.
}
\label{f6}
\end{figure}

%-----ACF
We calculate auto-correlation for each component of the node vectors $V_i^\alpha (t)$. The auto-correlation is defined as the Pearson correlation for a time series as a function of the time lag.  The auto-correlation function for the different components for different values of $\alpha$, and $i$ is shown 
in Fig~\ref{f4}. It is evident that the components do not have any significant auto correlation. 
%==========

We show the distributions for the elements of correlation tensors with different time windows $(2 \Delta T + 1)$ in Fig.~\ref{f5}~(a). It is observed that the peak of the distribution becomes sharper as we increase the time window and approaches a saturation form.  We also notice that if we consider the time window $(2 \Delta T + 1) < 5$, the distribution appears with two more peaks at $\pm 1$ due to high noise. The correlation tensor contains more noise as we decrease the values of the time window $(2 \Delta T + 1)$.        
The singular values  $ \rho_k^\gamma$ of correlation tensors with different time windows $(2 \Delta T + 1)$ is shown in Fig.~\ref{f5}~(b). We observe that the largest singular value decreases as the time window increases.    

To understand the significance of the correlation tensor, we need a reference correlation tensor. We consider the following null hypothesis for the reference correlation tensor. We assign uniform random numbers between $[-1,1]$ to the components of the embedding node vectors.
We then  calculate the randomized correlation tensor from these components of the embedding node vectors.  We compare the distributions for the elements of the empirical correlation tensor and randomized correlation tensor in Fig.~\ref{f6}~(a). We observe that the distribution for the randomized correlation tensor has a higher peak and it is symmetric around zero. The distribution for the empirical correlation tensor is asymmetric and has a positive mean. 

We further compare the singular values $ \rho_k^\gamma$  of the empirical and randomized correlation tensors in Fig.~\ref{f6}~(b). We observe that the largest singular value of the empirical correlation tensor lies significantly beyond the largest singular value of the randomized correlation tensor. 

\begin{figure}[tbh]
%\centering
\includegraphics[width=0.98\textwidth]{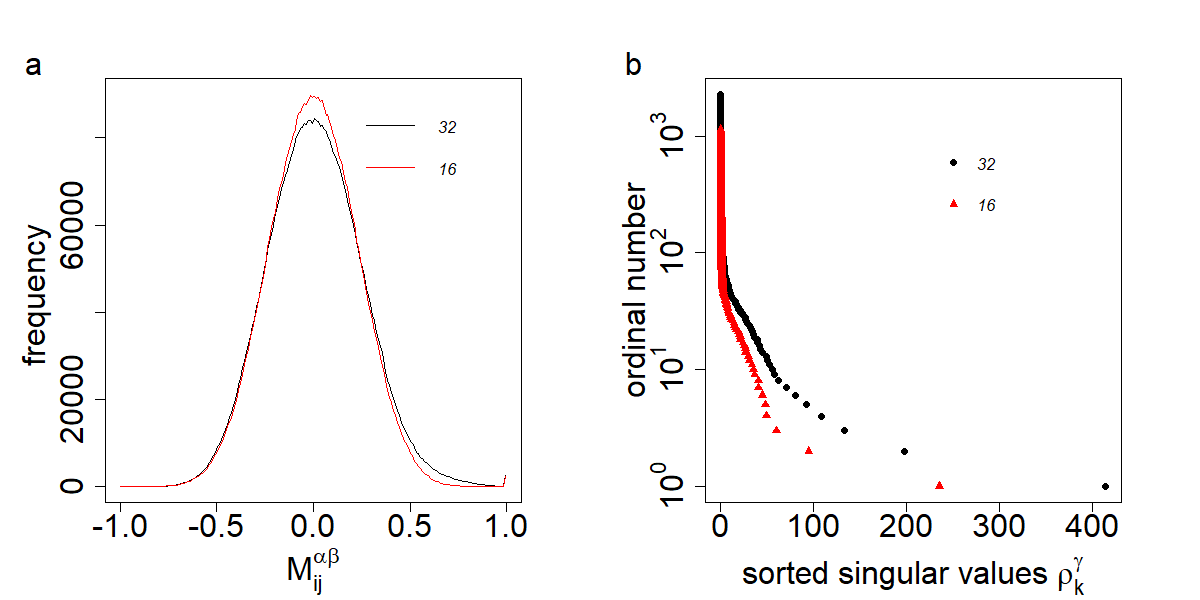}
\caption{ Correlation tensor with time window $2 \Delta T +1 = 21$ for embedding dimension $D = 16,$ and $32$, respectively (a) Distribution for the elements of correlation tensors.  (b) Sorted singular values for the corresponding correlation tensors. 
%You can embed figures using the \texttt{\textbackslash includegraphics} command. Basically, figures should appear where they are cited in the text. You do not need to separate figures from the main text when you use \LaTeX\ for preparing your manuscript.
}
\label{f7}
\end{figure}

We show the dependence of the correlation tensor on the dimension $D$ of embedding vector space in Fig.~\ref{f7}. The distribution of the elements of the correlation tensor becomes narrower as we increase the dimension, as shown in Fig.~\ref{f7}~(a). The corresponding singular values $ \rho_k^\gamma$ of the correlation tensors are shown in Fig.~\ref{f7}~(b). The largest singular value $ \rho_1^1$ increases as we embed the weekly networks in a higher dimension.

%\begin{figure}[tbh]
%\centering
%\includegraphics[width=0.98\textwidth]{Figures/spectra_D_v2.png}
%\caption{ Sorted singular values/D for corresponding correlation tensors with time window $2 \Delta T +1 = 21$ for embedding dimension $D = 16,$ and $32$ respectively . 
%You can embed figures using the \texttt{\textbackslash includegraphics} command. Basically, figures should appear where they are cited in the text. You do not need to separate figures from the main text when you use \LaTeX\ for preparing your manuscript.
%}
%\label{f8}
%\end{figure}

%Label figures, tables, and equations appropriately using the \verb|\label| command, and use the \verb|\ref| command to cite them in the text as ``\verb|as shown in Fig. \ref{f1}|". This automatically labels the numbers in numerical order.

%The \verb|minipage| environment can be used to place figures horizontally.

\section{Conclusion}
We have analyzed the XRP transactions focusing during the period between October, 02, 2017 and March 04, 2018. During this period, a significant price burst of XRP was observed. We have constructed weekly directed networks from XRP transactions. 
We have embedded these weekly networks in vector space, which capture the community structure information of the weekly networks in the embedded node vectors. The correlation tensor and its spectra are calculated using different time windows. 
It is observed that the the correlation tensor is sensitive to the time window. Lower the values of time window, it contains more noise.  As we increase the time window, the distribution of the elements of correlation tensor approaches to the fixed form.  
The results were compared with randomized counterpart. It is observed that the distribution for the elements of empirical correlation tensor differs significantly. The largest singular value of the empirical correlation tensor appears well above the largest
singular value of the randomized correlation tensor. 
Furthermore, we have presented the dependence of the  correlation tensor on model parameters in detail.  

This method is very useful and can capture valuable insights about the dynamical properties of the XRP transaction networks~\cite{chakraborty2022projecting}. The embedding of the nodes encodes the community structure in the node vectors. The correlation is calculated between the components of the regular node vectors. Therefore, the spectra of the correlation tensors can capture whether there is a change in the community structure of the XRP transaction networks.

\section*{Acknowledgements}
We thank the members of the Kyoto Univ.- RIKEN blockchain study group  for discussions.  YI acknowledges the grant, "University Blockchain Research Initiative", provided by Ripple, Inc. to Kyoto University for partial support to this work. 
% This work is partially supported by  Ripple Impact Fund 2022-247584 (5855).
%Use the \verb|\appendix| command if you need an appendix(es). The \verb|\section| command should follow even though there is no title for the appendix (see above in the source of this file).

%\bibliographystyle{apsrev4-1.bst}
%\bibliography{bibl}

\end{document}